\documentclass[aps,prb,twocolumn,showpacs]{revtex4}
\usepackage{graphicx}
\usepackage{amssymb}

\begin{document}

\draft

\title{Heat capacity of the quantum magnet TiOCl}

\author{J. Hemberger$^1$, M. Hoinkis$^2$, M. Klemm$^2$, M. Sing$^2$, R. Claessen$^2$, S. Horn$^2$,
and A. Loidl$^1$}

\address{%
$^1$Experimentalphysik V, Center for Electronic Correlations and
Magnetism, Universit\"at Augsburg, D-86135 Augsburg, Germany  \\%
$^2$ Experimentalphysik II, Institut f\"ur Physik, Universit\"at
Augsburg, D-86135 Augsburg, Germany
 }

\begin{abstract}
Measurements of the heat capacity $C(T,H)$ of the one-dimensional
quantum magnet TiOCl are presented for temperatures 2~K~$< T
<$~300~K and magnetic fields up to 5~T. Distinct anomalies at 91~K
and 67~K signal two subsequent phase transitions. The lower of
these transitions clearly is of first order and seems to be
related to the spin degrees of freedom. The transition at 92~K
probably involves the lattice and/or orbital moments. A detailed
analysis of the data reveals that the entropy change $\Delta S$
through both transitions is surprisingly small ($\sim 0.1 R$),
pointing to the existence strong fluctuations well into the
non-ordered high-temperature phase. No significant magnetic field
dependence was detected.
\end{abstract}



\maketitle

\section{Introduction}

The discovery of high-Tc superconductivity in the cuprates and of
colossal magneto-resistance in manganite perovskites generated an
enormous interest in transition-metal oxides (TMOs). TMOs are
characterized by an intimate coupling of spin, charge, orbital and
lattice degrees of freedom which is the origin of a number of
complex and exotic ground states. The discovery of a number of new
low-dimensional spin 1/2 quantum magnets was another result of
this renewed interest. After the first experimental observation of
a spin-Peierls scenario in the organic compound
TTF-CuBDT\cite{Bray75} more than 30 years ago, CuGeO$_3$
\cite{Haase93} was the first inorganic spin-Peierls system.
Another paramount example is NaV$_2$O$_5$, \cite{Isobe96} a spin
ladder with mixed-valent vanadium ions undergoing a charge
ordering transition \cite{Luedecke99}. Further interest in these
low-dimensional quantum magnets comes from the fact that upon
doping they may undergo metal-to-insulator transitions and
possibly reveal superconductivity. During the last decade {$\rm
TiOCl$} has been a distinguished candidate for a
resonating-valence bond ground state \cite{Beynon93} and has
emerged as a further quantum-spin magnet with a gapped ground
state \cite{Beynon93,Seidel03}.

Titanium-oxochloride has first been synthesized by Friedel and
Guerin in 1876. \cite{Friedel76} The first detailed report on
growth and characterization has been given by Sch\"afer et
al.~\cite{Schaefer58} almost 50 years ago. TiOCl crystallizes in
the orthorhombic FeOCl structure, consisting of Ti-O bilayers
separated by Cl layers. In this structure every Ti is surrounded
by 4 oxygen and 2 chlorine ions forming a distorted octahedron.
Originally, based on an analysis of susceptibility data combined
with band-structure calculations, it has been suggested that the
t$_{2g}$ orbitals are orbitally ordered, producing one-dimensional
antiferromagnetic $S = 1/2$ chains \cite{Seidel03}. The
high-temperature magnetic susceptibility could well be described
by a Bonner-Fisher type of behavior with an exchange constant $J =
660$~K. A sudden drop of the susceptibility almost to zero
indicated the opening of a spin gap at $T_{c1}=67$~K, while a
further anomaly has been detected at $T_{c2} \approx 95~K$.
\cite{Seidel03} Electron spin resonance \cite{Kataev03}, infrared
and Raman spectroscopy \cite{Caimi04,Lemmens04} as well as NMR
measurements \cite{Imai03} substantiated these results and
established a spin gap of the order of 430~K. The origin of the
gap has been identified as due to a first order spin-Peierls
transition at $T_{c1}$, as evidenced by the observation of Ti
dimerization in temperature-dependent x-ray diffraction.
\cite{Shaz04} The nature of a second transition at $T_{c2}$ is
still unclear. While a broadening of the NMR lineshape into a wide
continuum below $T_{c2}$ has been attributed to a possible
incommensurate (orbital) ordering \cite{Imai03}, no corresponding
superlattice reflections could be observed in x-ray scattering,
\cite{Shaz04} though there is some evidence for a structural
symmetry lowering from the high-temperature phase. Close to $T^*
\approx 135$~K the opening of a spin pseudo gap as detected in NMR
\cite{Imai03} and the onset of giant phonon anomalies in
infrared and Raman spectroscopy \cite{Lemmens04} indicated the
presence of strong spin fluctuations and a pronounced coupling to
the lattice, respectively, for temperatures up to $T^* \approx
135$~K.

Electronically, TiOCl is a Mott insulator with nominally trivalent
Ti. Electronic structure calculations using the $LDA+U$ predict a
Ti $3d_{xy}^1$ ground state, \cite{Seidel03,Saha04a} however,
coupling to optical phonon modes can lead to strong orbital
fluctuations within the $t_{2g}$ crystal field multiplet. The
importance of correlation effects has been elucidated in $LDA +
DMFT$ \cite{Craco04,Saha04b} studies.

In this 
brief experimental report  we provide detailed measurements of the
heat capacity of TiOCl to further unravel the nature of the two
phase transitions.

\section{Experimental Details}

Single crystals of TiOCl were prepared by chemical vapour
transport from TiCl$_3$ and TiO$_2$. \cite{Schaefer58} The samples
have been characterized using x-ray diffraction, SQUID and ESR
measurements at X-band frequencies.  The crystal symmetry was
found to be orthorhombic with lattice parameters of $a =
0.379$~nm, $b = 0.338$~nm and $c = 0.803$~nm. The magnetic
properties were found to be in excellent agreement with published
results \cite{Seidel03,Kataev03}. The heat capacity measurements
have been performed with a commercial PPMS from Quantum Design for
temperatures 1.8~K~$< T <$~300~K and in external magnetic fields
up to 5~T.

\section{Results and discussion}

\begin{figure}[tb]
\includegraphics[clip,angle=-0,width=85mm]{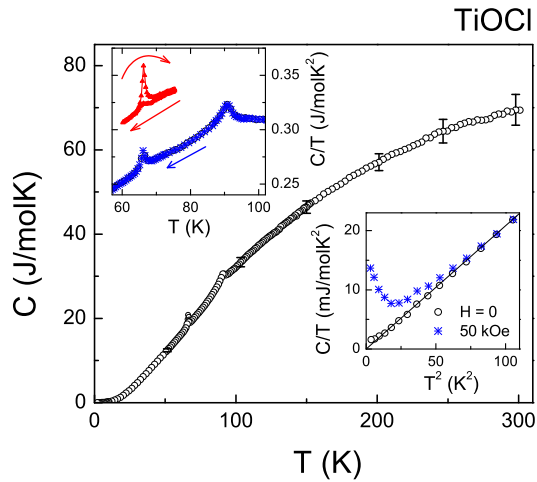}
\caption{Heat capacity of TiOCl vs. temperature. The vertical bars
indicate experimental uncertainties and reveal the scatter of
$C(T)$ in different measuring cycles with different temperature
stimuli and using different samples. Upper inset: Magnified region
of the two phase transitions. The heat capacity measured at 5~T
(stars) is compared to the heat capacity in zero
external field (circles), both measured on cooling. The
temperature region around the lower transition at $T_{c1}\approx
66.7$~K was remeasured (triangles) utilizing a reduced relaxation
amplitude of $\delta T \approx 0.5$~K on heating and subsequent
cooling. For clarity these data are shifted by 50~mJ/molK$^2$
Lower inset: Low-temperature heat capacity plotted as $C/T$ vs.\
$T^2$.  For the results in zero magnetic field a fit using a
Debye-derived phonon contribution is indicated as solid line.
\label{fig1}}
\end{figure}

Fig.~1 shows the central result of this investigation, the heat
capacity as function of temperature. The heat capacity is
characteristic for a three-dimensional solid with a Debye
temperature of order 200~K.
Superimposed we find two weak but distinct anomalies which are
clearly related to the two subsequent phase transitions at $T_{c1}
= 67$~K and $T_{c2} = 91$~K. In recent NMR experiments
\cite{Imai03} slightly different temperatures, namely 94~K and
66~K, respectively, have been determined. The effects of the phase
transitions on the specific heat are remarkably weak and are
almost lost under the large phonon-derived heat-capacity
contributions, and it is clear that the entropy covered by the two
anomalies is comparably small.

The upper inset in Fig.~1 shows the heat capacity (given as $C/T$
{\it vs.} $T$) in the temperature region of the phase transitions
for both zero field and an external magnetic field of 5~T measured
on cooling together with a special heating/cooling run in the
vicinity of $T_{c1}$ in zero-field, which will be discussed later.

Concerning the magnetic field dependence we observe no effects on
the heat capacity within experimental uncertainties. The moderate
magnetic fields used here neither shift the transition
temperatures, nor do they seem to affect the entropies involved in
the phase transitions. On the other hand, the small energy of our
fields of order 50~kOe$\times \mu_B$ ($\approx 0.06$~meV) is
negligibly small compared to the intrinsic magnetic energy scale
of TiOCl as estimated from $k_BT_{c1,2}$ (5.8 and 7.9~meV) or the
exchange constant $J$ (57~meV).

In order to analyze the nature of the detected anomalies in terms
of first or second order phase transitions, we performed
heating/cooling cycles across the phase transitions. As the
experimental setup utilizes a relaxation method, each data point
is related to the average over a temperature interval above the
initially stabilized temperature rather than to an exact
temperature.\cite{ppms} This means, that for the cooling sequence
the temperature is decreased between the data acquisition and
increased during the acquisition process itself. Thus, in the case
of hysteretic (i.e. first order) behavior, the actual transition,
together with the corresponding release of latent heat, only is
fully captured only in the heating branch of the measuring
sequence. Around $T_{c1}$ a significant difference between heating
and cooling can be detected, pointing towards a first order
transition (shifted curve in the inset of Fig.~1).
No such feature could be found for the upper transition at
$T_{c2}$.

The lower inset in Fig.~1 shows the low-temperature heat capacity
plotted as $C/T$ vs. $T^2$ to search for possible spin
contributions. In zero external magnetic field, $C(T)$ is fully
determined by phonon contributions and from a linear fit (solid
line) a Debye temperature of $\Theta_D' = 210$~K can be
determined. %
The low-temperature heat capacity becomes slightly enhanced in
magnetic fields of 5~T which can be explained by the increasing
importance of nuclear hyperfine contributions or the influence of
paramagnetic defects at the lowest temperatures.
For comparison, the low-temperature specific heat of the canonical
organic spin-Peierls compound (TMTTF)$_2$PF$_6$ \cite{tmttf}
displays besides the nuclear hyperfine and  $T^3$ phonon
contributions an additional quasi-linear ($\sim T^{1.2}$) term
which has been ascribed to low-energy excitations. A similar
contribution certainly can be excluded for TiOCl.

In the following analysis we will decompose the heat capacity into
phonon and spin ($S=1/2$) contributions, with the former
dominating the latter for all temperatures. Nevertheless, we are
able to perform a straight-forward analysis of the thermal
properties of TiOCl and we will show that we arrive at a rather
consistent description despite the large uncertainties mentioned
above.

\begin{figure}[b]
\includegraphics[clip,angle=-0,width=85mm]{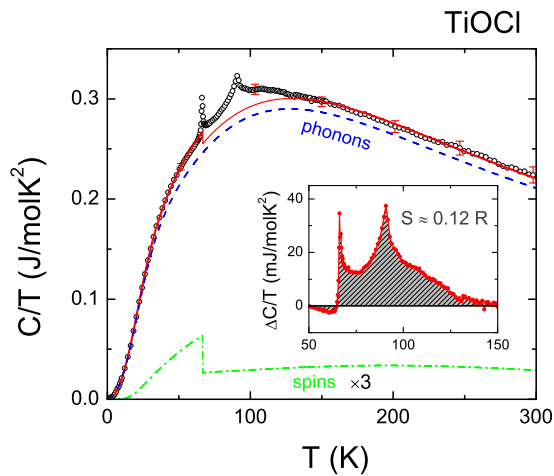}
\caption{Heat capacity of TiOCl plotted as $C/T$ vs.\ $T$. The
$S=1/2$ contribution including a mean-field spin-Peierls
transition as described in the text is indicated as dash-dotted
line. For representation purposes the calculated values have been
multiplied by a factor of 3. The fit to the phonon-derived
specific heat (see text) is indicated as dashed line. The sum of
both contribution is drawn as solid line and represents the best
fit to the experimental results. The inset shows the difference
between calculated and measured heat capacities in a limited
temperature regime.
\label{fig2}}
\end{figure}

Fig.~2 shows the heat capacity for the complete temperature regime
investigated, plotted as $C/T$ vs.\ $T$. Assuming a $S=1/2$ spin
chain the contribution of the spin chain for $T
> 67$~K is modelled by a Bonner-Fisher type of behavior \cite{Bonner64}
taking into account an exchange coupling $J = 660$~K. For the
low-temperature regime, {\it i.e.} below the spin-Peierls
transition at $T_{c1} = 67$~K, the expected exponential decrease
of the spin part of the heat capacity is simulated by a BCS-like
mean-field treatment\cite{Muehlschlegel59} with a temperature
dependent gap, $2\Delta_0 = 3.5$k$_BT_{c1}$, and, at $T_{c1}$, a
heat capacity jump of 1.43 times the contribution of the
spin-chain, as it was proposed for canonical spin-Peierls
systems.\cite{Bray75}
Using this approach we can calculate the spin-derived heat
capacity without any free parameter. The result (for
representation multiplied by a factor of 3) is shown as
dash-dotted line in Fig.~2. We are aware that this can only be a
very rough analysis, specifically having in mind that the actual
gap in TiOCl seems to be much larger and has recently been
determined as $2\Delta = 10\sim15 $k$_BT_{c1,c2}$ by NMR
experiments \cite{Imai03} and that the transition actually is of
first order. %
The disregard of these experimental facts gives an underestimation
of the jump height of the magnetic specific heat at the
spin-Peierls transition and a too broad decay below, leading
indeed to observable differences between the measured data and our
calculation closely below $T_{c1}$.
However, the overall entropy balance is not affected by these
shortcomings of our simple model. The total magnetic entropy at
high temperatures sums up to R$\ln2$ as expected for a spin 1/2
system.


The phonon system has been fitted assuming one Debye- and two
Einstein-type contributions, yielding a total of 5 free
parameters, namely, the mean Debye ($\Theta_D$) and two Einstein
temperatures ($\Theta_{E_{1,2}}$),
and the ratio of Debye to Einstein modes $R_{D/E}$,
as well as the ratio between the Einstein modes $R_{E_1/E_2}$. The
number of degrees of freedom ($N_f$), which of course should be
nine per formula unit of TiOCl, was kept fixed. The experimentally
determined heat-capacity values for 2~K~$< T < 65$~K and 130~K$< T
<$~300~K have been used for the fit of the heat capacity. For the
fitting procedure we included the spin contributions which have
been calculated parameter-free as outlined above. The total heat
capacity, spin and phonon contributions are shown as solid line in
Fig.~2. The lattice derived heat capacity is indicated as dashed
line. Despite this oversimplified model we arrive at an
astonishingly good description of the heat capacity over the
complete temperature range. The parameters as determined by the
best fit seem to be realistic: The characteristic temperatures,
$\Theta_D= 188$~K, $\Theta_{E_1} = 352$~K, and $\Theta_{E_2} =
614$~K, with the ratios $R_{D/E} = 0.30$ and $R_{E_1/E_2} = 0.51$
determining the relative weight of Debye and Einstein modes, are
reliable keeping in mind that the optical phonon modes in TiOCl
range from approximately 100~K to 700~K with dominant modes close
to our values for $\Theta_{E_{1,2}}$.\cite{Caimi04,Lemmens04} The
small discrepancy of the Debye temperature obtained from this
over-all fit to $C(T)$, compared to the value derived from the
low-temperature heat capacity only, can be explained by the
presence of relatively low lying optical modes.

Comparing the model heat capacity to the measured one in Fig.~2 we
find very good agreement for the smooth temperature evolution for
low ($T < T_{c1}$) and high ($T > T_{c2}$) temperatures. As
clearly seen, the main deviations arise in the temperature region
between the two phase transitions. %
The good agreement concerning the smooth part of the heat capacity
may seem surprising in the light of our simplified model. On the
other hand, the phonon part of our $C(T)$-model is reasonably
close to a realistic description involving the true phonon
spectrum and the spin part of the heat capacity bears only a small
weight. Thus, we believe that our model yields a reliable estimate
of the regular part of the specific heat, allowing us to separate
out the anomalous contribution due to the phase transitions at
$T_{c1}$ and $T_{c2}$. The inset of
Fig.~2 shows this extra heat capacity in a 
limited temperature range from 50 to 150~K. The integral over this
region gives an estimate of the entropy $\Delta S$ being released
when going through both transitions. $\Delta S$ is of the order of
0.12~R ($\pm 0.02$~R) vanishingly small compared to R$\ln(2)$
expected for a spin $S = 1/2$ system or an orbital doubly
degenerate state. The narrow peak at $T_{c1}$ corresponds to the
latent heat released at the phase transition where according to
magnetic measurements spin dimerisation appears. The larger
fraction of the entropy is covered under the phase transition at
$T_{c2}$ extending up to 135~K, exactly the temperature where in
NMR experiments the opening of a pseudo gap has been detected
\cite{Imai03}. Assuming that the phonon and spin derived heat
capacities are correctly described, we are left with the orbital
degrees of freedom only. However, if the orbitals undergo an
ordering transition at $T_{c2}$ the entropy again is much too low
and orbital fluctuations must extend to much higher temperatures,
even far above $T^*$, the characteristic temperature of the
opening of a pseudo gap. In this sense our results are compatible
with the strong fluctuation effects observed in NMR \cite{Imai03}
and in Raman scattering \cite{Lemmens04} and with the results of
density-functional calculations suggesting that TiOCl may be
subject to strong orbital fluctuations \cite{Saha04a}.
This indicates that the phase above $T_{c1}$ is
characterized by the presence of pronounced fluctuations retaining
some sort of short range order and thereby delaying the release of
the full entropy with temperature, as typical for one-dimensional
systems.
Unfortunately, our specific heat data do not allow us to decide on
the precise microscopic nature of the fluctuations (spin-Peierls
or orbital ordering).

 \section{Conclusions}

Summarizing, we have measured the heat capacity of TiOCl for
temperatures 2~K~$< T <$~300~K and magnetic fields up to 5~T. We
observed two anomalies at $T_{c1}=91$~K and $T_{c2}=67$~K
corresponding to two subsequent phase transitions.
The phase transition at 67~K, where strong spin-dimerization sets
in, reveals significant hysteresis effects and hence is of first
order. Most probably the orbital degrees of freedom are involved
in the phase transition at $T_{c2}$, which shows no thermal
hysteresis. The entropy below this anomaly is vanishingly small
indicating that the largest fraction of entropy is released at
considerable higher temperatures. The total heat capacity can
satisfactorily be explained describing the phonon spectrum with
Debye and Einstein modes and the magnetic excitations by a
Bonner-Fisher type heat capacity for $T > T_{c1}$ and the opening
of a spin gap below.
At low temperatures and zero external fields the heat capacity can
satisfactorily be described taking a phonon-derived $T^3$ term
into account only. No indication of contribution of low-energy
excitations can be detected.

\acknowledgments

We want to thank R.\ Valenti, J. Deisenhofer, and H.-A. Krug von
Nidda for helpful discussions. This work was partly supported by
the Bundesministerium f\"ur Bildung und Forschung (BMBF) via Grant
No.\ VDI/EKM 13N6917-A and by the Deutsche Forschungsgemeinschaft
(Sonderforschungsbereich 484 in Augsburg and project CL 124/3-3).


\bibliographystyle{prsty}

\end{document}